\documentclass[10pt,twocolumn,letterpaper]{article}

\usepackage{cvpr}              

\usepackage[dvipsnames]{xcolor}

\definecolor{cvprblue}{rgb}{0.21,0.49,0.74}
\usepackage[pagebackref,breaklinks,colorlinks,citecolor=cvprblue]{hyperref}

\title{LGFN: Lightweight Light Field Image Super-Resolution using Local Convolution Modulation and Global Attention Feature Extraction}

\author{  
	Zhongxin Yu,  Liang Chen\thanks{Corresponding author}, Zhiyun Zeng,   Kunping Yang, Shaofei Luo, \\Shaorui Chen,  Cheng Zhong 
	\\[1ex]
	Fujian Normal University \\ 
	{\tt\small wuyizhizi555@163.com \hspace{5pt} cl\_0827@126.com \hspace{5pt}  1304370458@qq.com \hspace{5pt}} \\
	{\tt\small kunpingyang@fjnu.edu.cn  \hspace{5pt}  \{shaofeiluo,1589177538,2998233739\}@qq.com \hspace{5pt}}
	}  

\begin{document}
\maketitle
\begin{abstract}
Capturing different intensity and directions of light rays at the same scene, Light field (LF) can encode the 3D scene cues into a 4D LF image, which has a wide range of applications (i.e., post-capture refocusing and depth sensing). LF image super-resolution (SR) aims to improve the image resolution limited by the performance of LF camera sensor. Although existing methods have achieved promising results, the practical application of these models is limited because they are not lightweight enough.
In this paper, we propose a lightweight model named LGFN, which integrates the local and global features of different views and the features of different channels for LF image SR. Specifically, owing to neighboring regions of the same pixel position in different sub-aperture images exhibit similar structural relationships, we design a lightweight CNN-based feature extraction module (namely, DGCE) to extract local features better through feature modulation. Meanwhile, as the position beyond the boundaries in the LF image presents a large disparity, we propose an efficient spatial attention module (namely, ESAM) which uses decomposable large-kernel convolution to obtain an enlarged receptive field and an efficient channel attention module (namely, ECAM). Compared with the existing LF image SR models with large parameter, our model has a parameter of 0.45M and a FLOPs of 19.33G, which has achieved a competitive effect. Extensive experiments with ablation studies demonstrate the effectiveness of our proposed method, which ranked the second place in the Track 2 Fidelity $\&$ Efficiency of NTIRE2024 Light Field Super Resolution Challenge and the seventh place in the Track 1 Fidelity.
\end{abstract}    
\section{Introduction}
\label{sec:intro}
LF cameras can capture varying intensities and directions of light rays within the same scene, encoding the 3D scene cues into a 4D LF image (comprising spatial and angular dimensions). This technology finds wide applications, including post-capture refocusing\cite{1vaish2004using,2wang2018selective}, depth sensing\cite{3shin2018epinet,4wang2022occlusion,5chao2023learning}, virtual reality\cite{6overbeck2018system,7yu2017light}, and view rendering\cite{8wu2021revisiting,9sitzmann2021light,10wang2022r2l,11attal2022learning}. However, due to the limitation of sensor performance, there exists a trade-off between the spatial resolution and angular resolution of LF images. How to improve the resolution of LF images is currently a prominent research challenge. 
\begin{figure}[t]
  \centering
   \includegraphics[width=0.47\textwidth]{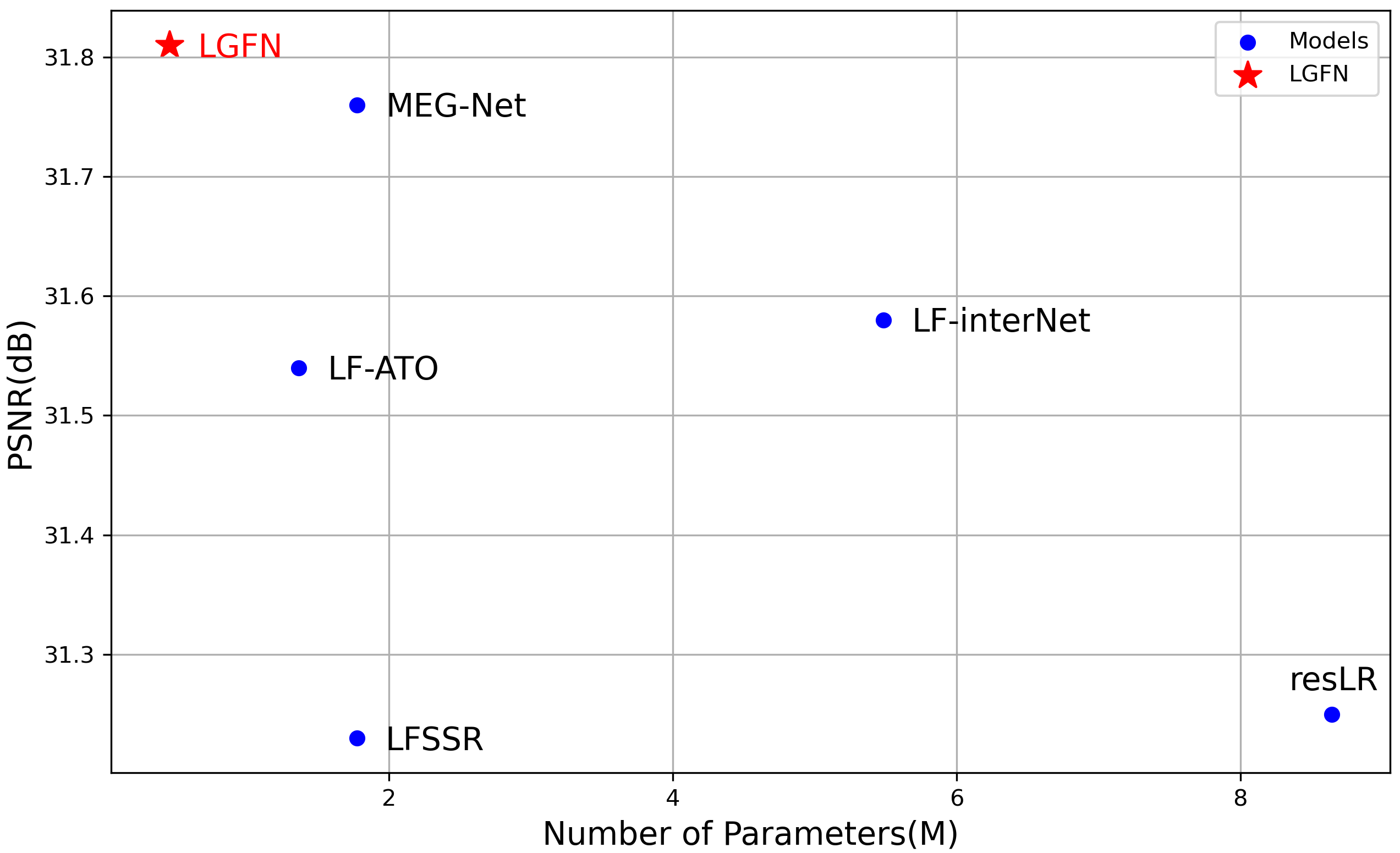}
   \caption{Comparisons of the parameters and PSNR of different LF image SR methods.}
   \label{fig:onecol}
\end{figure}

The traditional LF SR method\cite{12_v2_5bishop2011light,13_v2_6wanner2013variational,14_v2_7farrugia2017super,15_v2_8rossi2017graph,16_v2_9alain2017light} mainly focuses on how to find sub-pixel information and warp multi-view images based on estimated disparities. However, the performance of these methods heavily depends on accurate estimated disparities, which is difficult to achieve in low-resolution LF images and complex imaging environments such as occlusion and non-Lambert reflection\cite{17_v2_10meng2019high}.

In recent years, deep learning-based methods have been widely used. Yoon et al.\cite{18_v2_11yoon2015learning} proposed the first CNN-based LF image SR model (i.e., LFCNN), which used SRCNN to super-resolve each sub-aperture image (SAI). Afterwards, many methods have adopted the CNN-based methods to integrate different angle information to improve the performance of SR\cite{19_v2_12wang2018lfnet,20_v2_13zhang2019residual,21_v2_14cheng2019light,22_v2_15wang2020light,23_v2_16zhang2021end}. Besides directly processing the 4D LF data, some methods extracted two kinds of features by designing spatial and angular feature extractor, and interacted with each other\cite{24_v2_17yeung2018light,25_v2_18wang2020spatial,26_v2_19wang2022detail,27_v2_20liu2021intra,28_v2_21wang2022disentangling}. Apart from the CNN-based LF image SR methods, Transformer-based LF methods have also been proposed. Wang et al.\cite{26_v2_19wang2022detail} proposed a detail-preserving Transformer (DPT) for LF image SR. Liang et al.\cite{29_v2_22liang2022light} proposed a simple yet efficient Transformer method for LF image SR. Liang et al.\cite{30_v2_23liang2023learning} proposed EPIT to LF image SR by learning non-local space and angle cooperation. Jin et al.\cite{31_v2_24jin2023distgepit} proposed DistgEPIT model that learn global features and local features of LF images by designing an attention branch and a convolution branch respectively.	While existing methods have achieved promising results, the practical application of these models is limited due to excessive parameters and FLOPs. As shown in Fig.1, the parameters of some existing LF image SR methods are mostly above 1M.  This limitation prompts our research into lightweight LF image SR.

\begin{figure}[t]
  \flushleft 
   \includegraphics[width=0.50\textwidth]{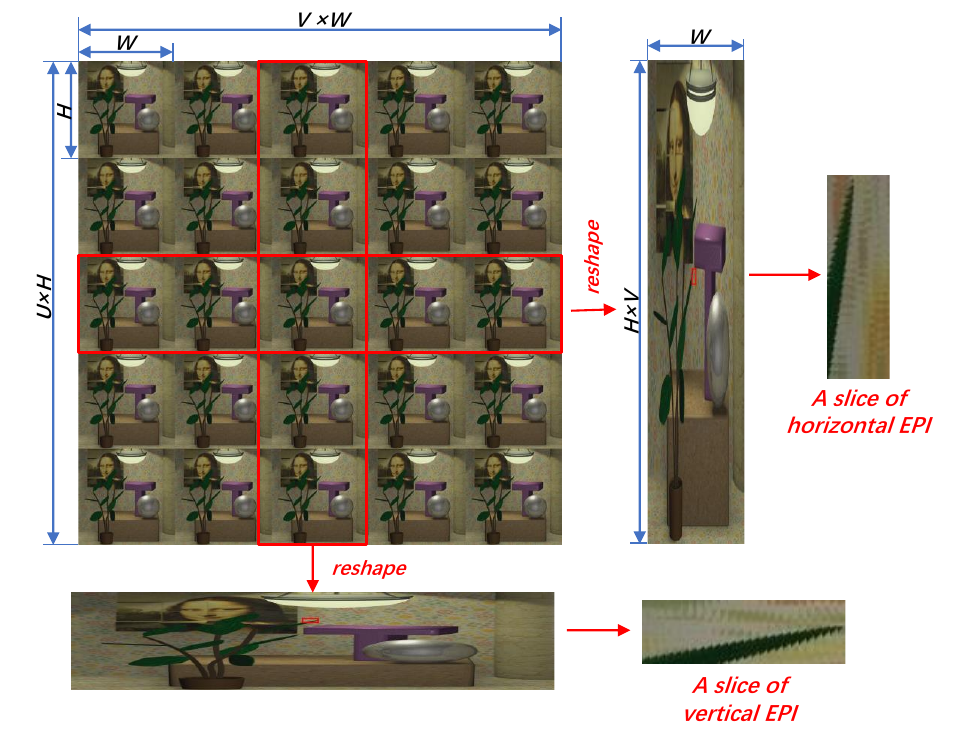}
   \caption{The epipolar plane images(EPI) sample of 4D LF is acquired with fixed angular coordinate and a fixed spatial coordinate. Specifically, the horizontal EPI is obtained with constants u and h, and the vertical EPI is obtained with constant v and w. On the one hand, the EPIs capture spatial structures such as edges or textures, and the adjacent areas corresponding to the same pixel position across different SAIs exhibit similar structural relationships. On the other hand, the EPIs reflect the disparity information via line patterns of different slopes, whereas positions located outside the boundary in the LF image exhibit a large parallax.}
      \label{fig:onecol2} 
\end{figure}

As shown in Fig.2, adjacent areas at the same pixel position across different SAIs exhibit similar structural relations, which are suitable for processing by the local feature extraction module. On the other hand, the position outside the boundary in the LF image exhibits a large parallax, which requires aggregation of contextual features across SAIs for processing.

To consider these two aspects and the requirement of lightweight model, we propose a lightweight local and global feature learning model named LGFN. By integrating both local and global features of different views and the features of different channels, our lightweight model can achieve competitive results compared with the existing model with larger parameters. Specifically, our model chooses the convolution module with local representation. Different from the existing CNN-based methods\cite{19_v2_12wang2018lfnet,20_v2_13zhang2019residual,21_v2_14cheng2019light,22_v2_15wang2020light,23_v2_16zhang2021end} which use complex network structure, we propose a simple yet efficient convolution module designed to extract local features through feature modulation performed by two parallel convolution branches.

In addition, we choose attention mechanism to extract contextual features.Different from the existing Transformer-based methods \cite{26_v2_19wang2022detail,29_v2_22liang2022light,30_v2_23liang2023learning,31_v2_24jin2023distgepit}with quadratic complexity over the number of visual tokens, we propose a simple yet efficient spatial attention module, whose attention weight branch uses decomposable 
large-kernel convolution to obtain an enlarged receptive field, and multiplies it with identity branch to extract contextual features. Besides, an efficient channel attention module (namely, ECAM) is introduced to enhance the features between channels. In order to further refine the feature extraction, we extract the local and global features along the horizontal and vertical directions respectively.

Our main contribution can be summarized as:
{\begin{enumerate} 
\item We design a lightweight convolution modulation module named DGCE to extract the local spatial features of LF images. A lightweight spatial attention module named ESAM with enlarged receptive field is designed to extract global features. In order to further refine the feature extraction, we extract the local and global features along the horizontal and vertical directions respectively.

\item We design an efficient channel attention module named ECAM and use the statistical information of channel direction to model the correlation between different channels.

\item We propose a light-weight LF image SR model named LGFN, which has a parameter of 0.45M and a FLOPs of 19.33G. Compared with the existing LF image SR models with large parameter, it has achieved a competitive effect, and won the second place in the Track 2 Fidelity \& Efficiency of NTIRE2024 Light Field Super Resolution Challenge and the seventh place in the Track 1 Fidelity.
\end{enumerate}} 

\section{Related Work}
\label{sec:formatting}

LR image SR methods can be divided into traditional non-learning methods, CNN-based methods and Transformer-based methods.

\subsection{Traditional Methods}

The traditional LF image SR methods mainly focuses on how to find sub-pixel information and warp multi-view images based on estimated disparities. Based on estimated disparities, Bishop et al.\cite{12_v2_5bishop2011light} used a Bayesian deconvolution method to super-resolve LF images. Wanner et al.\cite{13_v2_6wanner2013variational} used EPI to estimated disparity maps and proposed variational framework for LF image SR. Farrugia et al.\cite{14_v2_7farrugia2017super} proposed an example-based LF image SR method, enhancing spatial resolution consistently across SAIs through learning linear projections from reduced-dimension subspaces and angular super-resolution via multivariate ridge regression. Besides, optimization-based methods have also been proposed. Alain et al.\cite{16_v2_9alain2017light} adopted an optimization method to solve ill-posed LF image SR problem based on sparsity prior. Rossi et al.\cite{15_v2_8rossi2017graph} coupled the multi-frame information with a graph regularization, adopted convex optimization method to solve LF image SR problem. 

However, the performance of these methods depends heavily on accurate estimated disparities, it is difficult to achieve in low-resolution LF images and complex imaging environments such as non-Lambertian surfaces or occlusions\cite{17_v2_10meng2019high}.

\subsection{CNN-based Methods}
In recent years, deep learning-based method have been widely used. Yoon et al. \cite{18_v2_11yoon2015learning} proposed the first CNN-based LF image SR model (i.e., LFCNN), which used SRCNN to super-resolve each SAI. Similarly, Yuan et al.\cite{32_v2_25yuan2018light} uses EDSR to super-resolve each SAI. Afterwards, many methods have adopted the CNN-based methods to integrate different angle information to improve the performance of SR. Wang et al.\cite{19_v2_12wang2018lfnet}proposed a bidirectional recurrent CNN network iteratively model spatial relations between horizontally or vertically adjacent SAIs. Zhang et al.\cite{20_v2_13zhang2019residual} proposed resLF network that used four-branch residual network extracted features from SAI images along four different angular directions. Zhang et al.\cite{23_v2_16zhang2021end} proposed a 3D convolutions network extracted features from SAI images along different angular directions. Cheng et al. \cite{21_v2_14cheng2019light}considered the characteristics of internal similarity and external similarity of LR images, and fused these two complementary features for LF image SR. Meng et al. \cite{17_v2_10meng2019high} directly used 4D convolution to extract the angle information and spatial information of the LF image. Wang et al.\cite{22_v2_15wang2020light}designed an angular deformable alignment module (ADAM) for feature-level alignment, and proposed a collect-and-distribute approach to perform bidirectional alignment between the center-view feature and each side-view feature. In addition to directly processing the 4D LF data, some methods disentangled the 4D LFs into different subspaces for SR. Wang et al. \cite{25_v2_18wang2020spatial} proposed a spatial and angular feature extractor to extract the corresponding spatial and angular information from the MacPI image, and proposed LF-InterNet\cite{27_v2_20liu2021intra}and DistgSSR\cite{28_v2_21wang2022disentangling}to repetitively interact the two features. 

Besides the aforementioned  methods to improve SR performance, some methods try to solve the complex degradation problem facing the real world. To address the issue of the domain gap in LF image SR, Cheng et al.\cite{36_replace_cheng2021light} proposed a 'zero-shot' learning framework. They divided the end-to-end model training task into three sub-tasks: pre-upsampling, view alignment, and multi-view aggregation, and subsequently tackled each of these tasks separately by using simple yet efficient CNN networks.
Xiao et al.\cite{34_v2_39xiao2023toward} proposed the first real-world LF image SR dataset called LytroZoom, and proposed an omni-frequency projection network(OFPNet), which deals with the spatially variant degradation by dividing features into different frequency components and iteratively enhancing them. Wang et al.\cite{33_v2_38wang2024real} developed a LF degradation model based on the camera imaging process, and proposed LF-DMnet that can modulate the degradation priors into CNN-based SR process.
\subsection{Transformer-based Methods}

In addition to the CNN-based LF image SR methods, Transformer-based LF methods have also been proposed. Wang et al.\cite{26_v2_19wang2022detail} proposed a detail-preserving Transformer (DPT) for LF image SR, which regards SAIs of each vertical or horizontal angular view as a sequence, and establishes long-range geometric dependencies within each sequence via a spatial-angular locally-enhanced self-attention layer. Liang et al. \cite{29_v2_22liang2022light}proposed a simple yet efficient Transformer method for LF image SR, in which an angular Transformer is designed to incorporate complementary information among different views, and a spatial Transformer is developed to capture both local and long-range dependencies within each SAI. By designing three granularity aggregation units to learn LF feature, Wang et al.\cite{35_v2_26wang2022multi}proposed a multi-granularity aggregation Transformer (MAT) for LF image SR; Liang et al. \cite{30_v2_23liang2023learning}  proposed EPIT to LF image SR by learning non-local space and angle cooperation. Jin et al. \cite{31_v2_24jin2023distgepit} proposed DistgEPIT model that learns global features and local features of LF images by designing an attention branch and a convolution branch respectively.

Although the existing models have achieved promising results, their model parameters and FLOPs are not lightweight enough, which limits their practical application. In order to solve these problems, we propose a lightweight SR model of LF image by designing efficient modules.
\section{Method}

As mentioned above, the LF image SR needs to consider the local similarity between SAI subgraphs on the one hand, and the disparity problem behind different subgraphs on the other hand, which urges us to consider the methods of local and global feature extraction.

In order to design a lightweight model with fewer parameters and FLOPs, we choose to reduce the high-dimensional feature space to the low-dimensional feature subspace, and design an efficient local and global feature extraction model to achieve LF image SR.

 \begin{figure*}[t]
 \centering
 \includegraphics[width=17.8cm]{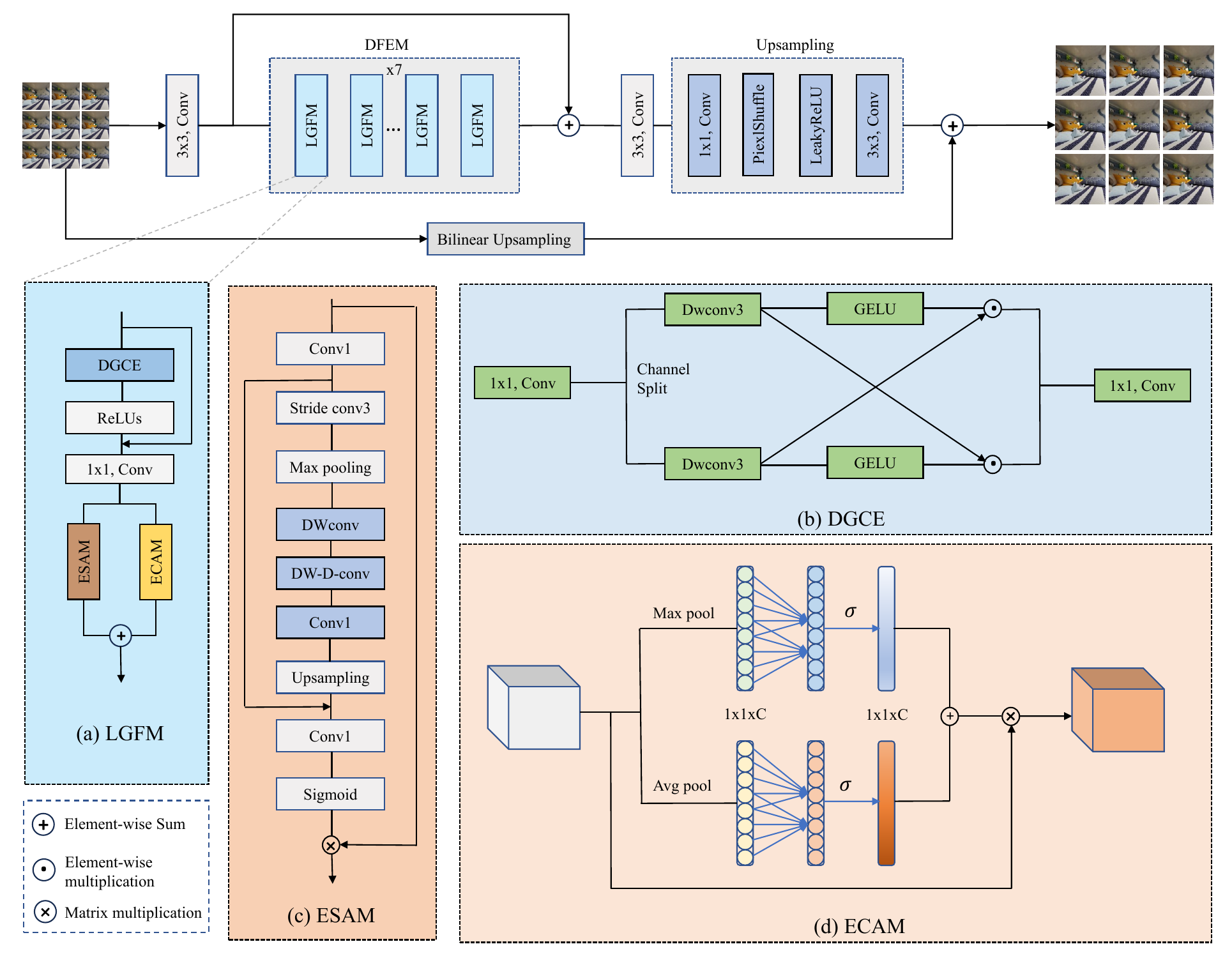}
 \caption{An overview of our LGFN network. (a) Local and global deep feature extraction module (LGFM); (b) Double-gated convolution extraction module (DGCE); (c) Efficient spatial attention module (ESAM); (d) Efficient channel attention module (ECAM). Given SAIs as inputs, we adopt bilinear upsamping to initial content of the original images. For feature extraction, we first use a 3D convolution to extract shallow features, then use the deep feature extraction module to get them, and finally use the upsampling module to obtain ultimate super-resolved SAI results. The depth feature extraction module (DFEM) includes seven local and global feature extraction modules, which are composed of DGCE, ESAM and ECAM.}\label{fig:overnetwork}
 \end{figure*}

\subsection{Network Architecture}

Specifically, as illustrated in Fig.3, our LF image SR model mainly consists of three components: shallow feature extraction, deep feature extraction and up-sampling module. Given an input LF low-resolution image $F_{LR}\in R^{U\times V\times H\times W}$ denote an LR SAI array with $U \times V $ SAIs of resolution $ H\times W $. Our method takes $ F_{LR}$ as its input and generates a HR SAI array of size $ F_{HR}\in R^{U\times V\times s H\times s W} $,where $s$ denotes the upsampling factor. 

Firstly, in the shallow feature extraction part, the low-resolution 4D LF image is upsampled using bilinear  interpolation to the size of $sH\times sW$. Meanwhile, it is converted to $F_0\in R^{1\times U V\times H\times W}$format and passed through a 1×3×3 spatial convolution to extract the shallow feature $F_{init}$, and the number of channels is increased from 1 to 64:
\begin{equation}  
F_{init}=H_{conv}(F_{LR})
\end{equation}
where  $H_{conv}(.)$  denotes 3D convolution operation.

Next, the shallow features $F_{init}$ pass through the jump connection and the deep feature extraction module(DFEM) respectively to obtain the jump connection feature and the deep feature, and they are fused by a 3D convolution process.
\begin{equation}  
F_{1}=H_{DFEM}(F_{init})+F_{init}
\end{equation}
\begin{equation}  
F_{fuse}=H_{conv}(F_{1})
\end{equation}
where $H_{DFEM}(.)$ and $H_{conv}(.)$  denote deep feature extraction module and 3D convolution operation, respectively.

Finally, the fused features $F_{fuse}$ pass through an up-sampling module consisting of 1×1 convolution, piexlshuffle, LeakyReLU and 3×3 convolution. In addition, the final restored image  $F_{HR}$ is obtained by adding the initial features after bilinear interpolation:
\begin{equation}  
F_{HR}=H_{upsampling}(F_{fuse})+H_{bilinear}(F_{LR})
\end{equation}
where $H_{upsamping}(.)$ denotes up-sampling module, and $H_{bilinear}(.)$ denotes bilinear interpolation.

\subsection{Local and Global Deep Feature Extraction}
DFEM includes seven local and global feature extraction modules (LGFM). The LGFM consists of three components: double-gated convolution extraction module (DGCE), efficient spatial attention module (ESAM) and efficient channel attention module (ECAM), as shown in Fig.3(a).	

 \subsubsection{Double-Gated Convolution Extraction Module} Owing to neighboring regions of the same pixel position in different SAIs exhibit similar structural relationships, which is suitable for processing with local feature extraction module. 

 Some studie\cite{37_v2_40guo2023visual,38_v2_41yang2022focal,41_v2_44ma2024efficient}indicate that modulation mechanism provides satisfactory performance and is theoretically efficient (in terms of parameters and FLOPs).  Therefore, we design a local feature extraction module based on feature modulation, as shown in Fig.3(b). In order to extract the local features better, the shallow features first undergo a 1×1 convolution, and then are cut into two halves along the channel. One half of the features undergoes a 3×3 depth-wise convolution and GELU function, and the other half of the features undergoes pixel-wise multiplication with the corresponding pixels to obtain the enhanced local features. After they are added to each other, they are fused by a 1×1 convolution:
\begin{equation}  
F_{21},F_{22}=Split(H_{conv1}(F_{init}))
\end{equation}
\begin{equation}  
F_{23}= \Phi(H_{dwconv3}(F_{21}))\odot F_{22}+\Phi(H_{dwconv3}(F_{22}))\odot F_{21}
\end{equation}
\begin{equation}  
F_{DGCE}=H_{conv1}(F_{23})
\end{equation}
\begin{equation}  
F_{24}=H_{conv1}(F_{init}+F_{DGCE})
\end{equation}
where $\Phi(.)$ denotes activation function GELU(.), $\odot$ denotes element-wise product, $Split(.)$ denotes split features along the channel, $H_{conv1}(.)$ and $H_{dwconv3}(.)$ denote 1×1 convolution and 3×3 depth-wise convolution respectively.

 \subsubsection{Efficient Spatial Attention Module}
 Owing to the position beyond the boundaries in the LF image presents a large disparity, which requires aggregate context features among different SAIs, therefore we propose a simple yet efficient spatial attention module, as shown in Fig.3(c). In order to reduce FLOPs, a 1×1 convolution is used to reduce the number of channels, and then strided convolution and max pooling are used to further reduce the height and width of features. In order to further increase the receptive field of spatial attention, the large-kernel convolution is decomposed into a depth-wise convolution\cite{37_v2_40guo2023visual}, a dilated convolution and a 1×1 point convolution, which can capture long-range relationships while maintaining low computational cost and few parameters. Then, the spatial resolution is restored to the original scale by up-sampling, and the number of channels is restored to the original number by a convolution. Therefore, attention with a large receptive field is obtained, which is convenient for the next attention calculation. The difference between our ESAM and other spatial attention modules is that the receptive field has been enlarged.
\begin{equation}  
F_{25}=H_{conv1}(F_{24})
\end{equation}
\begin{equation}  
F_{26}=H_{Maxpool}(H_{stride}(F_{25}))
\end{equation}
\begin{equation}  
F_{27}=H_{unsampling}(H_{LKA}(F_{26}))
\end{equation}
\begin{equation}  
F_{28}=H_{conv1}(F_{25}+F_{27})
\end{equation}
\begin{equation}  
F_{29}=sigmoid(F_{28})\otimes F_{24}
\end{equation}
where $H_{LKA}(.)$ denotes decomposable large-kernel convolution operation.

 \begin{figure*}[t]
 \centering
 \includegraphics[width=17.6cm]{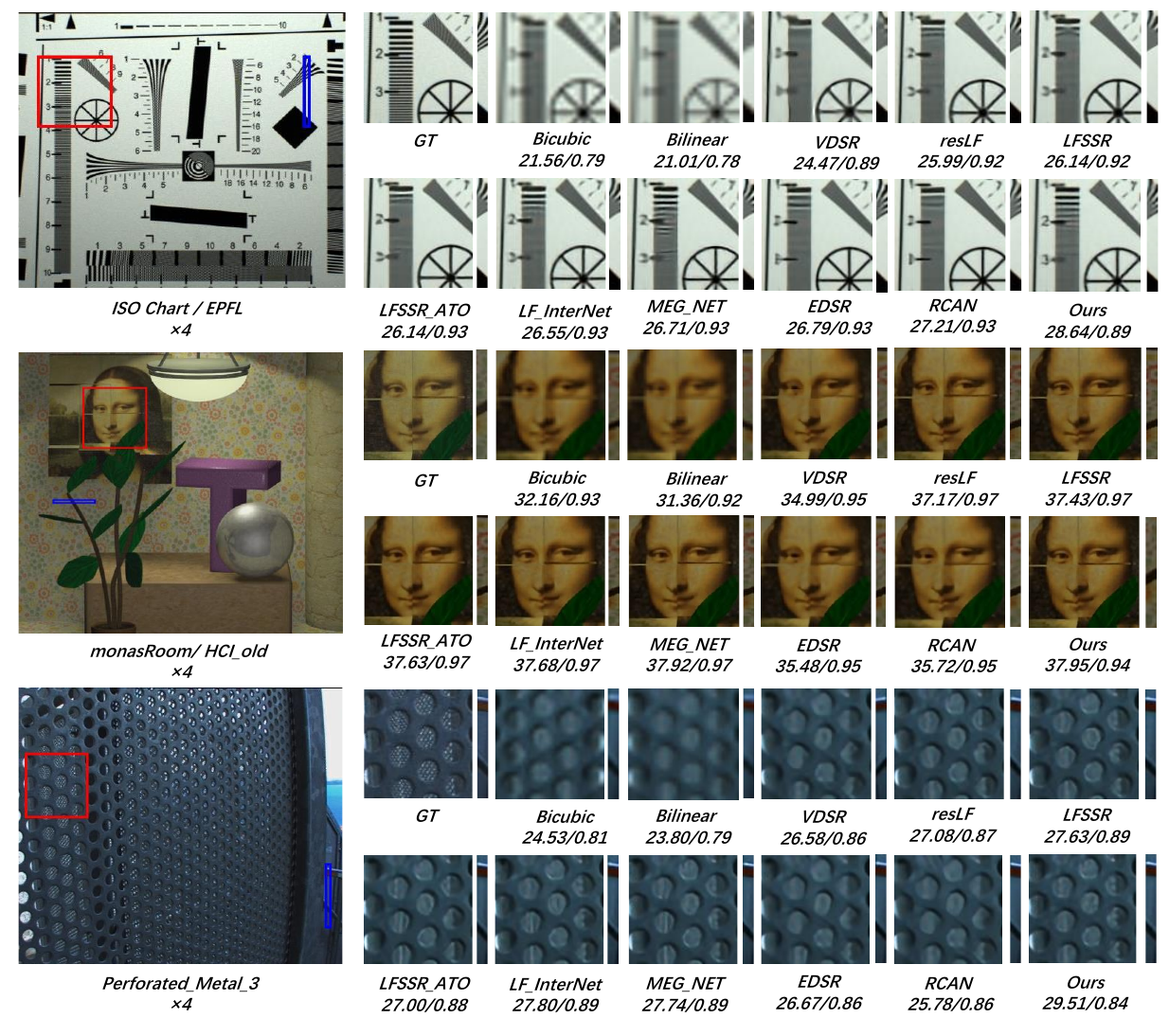}
 \vspace{-10pt}
 \caption{Qualitative results for 4x SR. The super-resolved center view images are presented for detailed texture comparison. The corresponding PSNR/SSIM scores of different methods on the presented scenes are also reported below.}
 \label{fig:overnetwork2} 
  \vspace{-10pt}
 \end{figure*}

 \subsubsection{Efficient Channel Attention Module}
Some studies\cite{20_v2_13zhang2019residual,22_v2_15wang2020light,25_v2_18wang2020spatial}show that the channel-wise features can improve LF image SR. After ESAM, we further design an efficient channel attention, as shown in Fig.3(d). For the input features, after adaptive maximum pooling and adaptive average pooling, each channel and its three adjacent channels are convolved with convolution kernel of 3 to capture local cross-channel interaction information, and two types of channel attention are obtained by sigmoid function, and then the channel attention is calculated after adding them:
\begin{equation}  
F_{30}=H_{conv3}(H_{Maxpool}(F_{29})
\end{equation}
\begin{equation}  
F_{31}=H_{conv3}(H_{Avgpool}(F_{29})
\end{equation}
\begin{equation}  
F_{32}=(sigmoid(F_{30})+sigmoid(F_{31}))\otimes F_{29}
\end{equation}

In order to further refine the feature extraction, LGFM extract the local and global features along the horizontal and vertical directions respectively.

\section{Experiments}

In this section, we first describe the experimental details, and then carry out specific control experiments and ablation experiments.

\subsection{Datasets and Implementation Details}

We used the five public LF datasets: EPFL\cite{44_replace_rerabek2016new}, HCInew \cite{42_v2_29honauer2017dataset}, HCIold\cite{43_v2_30wanner2013datasets}, INRIA\cite{39_v2_42le2018light}, and STFgantry\cite{40_v2_43vaish2008new}, following the same training and testing partition as in\cite{22_v2_15wang2020light}.

\begin{table*}[htbp] 
\centering 
\small
\caption{Overall PSNR/SSIM metrics comparison among the other prestigious approaches for 4 x SR. The best results are in red, the second best in black.} 
\begin{tabular}{c|c|ccccc|c} 
\toprule 
Methods & \#Param & EPFL & HCInew & HCIold & INRIA & STFganry & Average \\  

\hline  
Bilinear & - & 24.57 / 0.8158 & 27.09 / 0.8397 & 31.69 / 0.9256 & 26.23 / 0.8757 & 25.20 / 0.8261 & 26.95 / 0.8566 \\  
Bicubic & - & 25.14 / 0.8324 & 27.61 / 0.8517 & 32.42 / 0.9344 & 26.82 / 0.8867 & 25.93 / 0.8452 & 27.58 / 0.8701 \\  
VDSR\cite{45_v2_32kim2016accurate} & 0.665M & 27.25 / 0.8777 & 29.31 / 0.8823 & 34.81 / 0.9515 & 29.19 / 0.9204 & 28.51 / 0.9009 & 29.81 / 0.9066 \\ 
EDSR \cite{46_v2_33lim2017enhanced} & 38.89M & 27.84 / 0.8854 & 29.60 / 0.8869 & 35.18 / 0.9536 & 29.66 / 0.9257 & 28.70 / 0.9072 & 30.20 / 0.9118 \\
RCAN \cite{47_v2_34zhang2018image} & 15.36M & 27.88 / 0.8863 & 29.63 / 0.8886 & 35.20 / 0.9548 & 29.76 / 0.9276 & 28.90 / 0.9131 & 30.27 / 0.9141 \\

\hline  
resLF \cite{20_v2_13zhang2019residual} & 8.646M & 28.27 / 0.9035 & 30.73 / 0.9107 & 36.71 / 0.9682 & 30.34 / 0.9412 & 30.19 / 0.9372 & 31.25 / 0.9322 \\ 
LFSSR \cite{24_v2_17yeung2018light} & 1.774M & 28.27 / 0.9118 & 30.72 / 0.9145 & 36.70 / 0.9696 & 30.31 / 0.9467 & 30.15 / 0.9426 & 31.23 / 0.9370 \\ 
LF-ATO \cite{49_v2_36jin2020light} & 1.364M & 28.52 / 0.9115 & 30.88 / 0.9135 & 37.00 / 0.9699 & 30.71 / 0.9484 & 30.61 / \textbf{\color{black}0.9430} & 31.54 / 0.9373 \\ 
LF-InterNet\cite{25_v2_18wang2020spatial} & 5.483M & 28.67 / \textbf{\color{red}0.9162} & \textbf{\color{black}30.98} / \textbf{\color{black}0.9161} & \textbf{\color{black}37.11 / 0.9716} & 30.61 / \textbf{\color{red}0.9491} & \textbf{\color{black}30.53} / 0.9409 & 31.58 / \textbf{\color{black}0.9388} \\ 
MEG-Net\cite{23_v2_16zhang2021end} & 1.775M & 28.74 / \textbf{\color{black}0.9160} & \textbf{\color{red}31.10 / 0.9177} & \textbf{\color{red}37.27 / 0.9716} & 30.66 / \textbf{\color{black}0.9490} & \textbf{\color{red}30.77 / 0.9453} & 31.71 / \textbf{\color{red}0.9399} \\  

\hline  
LGFN-C & 0.45M & \textbf{\color{red}30.18} / 0.8698 & 30.42 / 0.8370 & 36.31 / 0.9283 & \textbf{\color{black}32.05} / 0.9040 & 30.05 / 0.9214 & \textbf{\color{black}31.80} / 0.8921 \\  
LGFN-P & 0.45M & \textbf{\color{black}30.05} / 0.8677 & 30.51 / 0.8681 & 36.29 / 0.9282 & \textbf{\color{red}32.08} / 0.9037 & 30.11 / 0.9207 & \textbf{\color{red}31.81} / 0.8977 \\  
\bottomrule 
\end{tabular}    

\label{tab:my_table}  
\end{table*}  

\begin{table*}[htbp] 
\centering 
\small
\caption{Ablation experiments operated on 4x SSR task. Note that the mode in the table refers to the connection mode of ESAM and ECAM modules.} 
\begin{tabular}{c|c|c c c|c c c c c |c|c}  
\toprule 
Mode & \#Param & DGCE & ESAM & ECAM & EPFL & HCLnew & HCLold & INRIA & STFgantry & Average & PSNR \\ 

\hline  
Parallel & 453.6k & \checkmark & \checkmark & \checkmark & 30.0461 & 30.5145 & 36.2853 & 30.0823 & 30.1098 & 31.8076 & Baseline \\
Cascade & 453.6k & \checkmark & \checkmark & \checkmark & 30.1782 & 30.4169 & 36.3102 & 32.0481 & 30.0533 & 31.8014 & -0.0062 \\ 
- & 452.9k & \checkmark & \checkmark & $\times$ & 29.8148 & 30.5509 & 36.4470 & 31.7838 & 30.3020 & 31.7804 & -0.0272 \\ 
- & 409.5k & \checkmark & $\times$ & \checkmark & 29.8481 & 30.5490 & 36.1630 & 31.7874 & 30.1549 & 31.6995 & -0.1081 \\ 
- & 409.5k & \checkmark & $\times$ & $\times$ & 29.8156 & 30.3048 & 36.0778 & 32.1215 & 29.9114 & 31.6462 & -0.1614 \\ 
Parallel & 147.0k & $\times$ & \checkmark & \checkmark & 27.4510 & 27.9710 & 32.7504 & 28.9564 & 26.7003 & 28.7796 & -3.0280 \\
\bottomrule 
\end{tabular}    

\label{tab:my_table2}  
\end{table*}

 {\bf Data Augmentation.} All LFs in the released datasets used the bicubic downsampling approach to generate LF patches of size 32×32. We performed random horizontal flipping, vertical flipping, and 90-degree rotation to augment the training data by 8 times. Note that, the spatial       and angular dimension need to be flipped or rotated jointly to maintain LF structures.

 {\bf Regularization.} Our network was trained using the L1 loss and FFT Charbonnier loss with weights of 0.01 and 1 respectively. Optimized using the Adam method with $\beta$1 = 0.9, $\beta$2 = 0.999 and a batch size of 1. Our model was implemented in PyTorch on a PC with a NVidia RTX 3060 GPU. The learning rate was initially set to 2x10-4 and decreased by a factor of 0.5 for every 15 epochs. The training was stopped after 100 epochs.

We used the PSNR and SSIM computed only on the Y channel of images as quantitative metrics for performance evaluation. To compute the metric scores for a dataset containing M scenes, we firstly computed the average score of each scene by separately averaging the scores over all SAIs. Then metric score for the dataset is determined by averaging the scores over the M scenes.

\subsection{Comparison to state-of-the-art methods}

We compared LGFN to several state-of-the-art methods, including five SISR methods: Bilinear, Bicubic, VDSR \cite{45_v2_32kim2016accurate}, EDSR\cite{46_v2_33lim2017enhanced}, RCAN\cite{47_v2_34zhang2018image} and other five recent LF image SR methods: resLF\cite{48_replaceNTIRE2024-LFSR}, LFSSR \cite{32_v2_25yuan2018light}, LF-ATO\cite{49_v2_36jin2020light}, LF-InterNet\cite{25_v2_18wang2020spatial}, and MEG\_Net\cite{23_v2_16zhang2021end}.

 {\bf Quantitative Results.} As shown in Table 1, compared with other models with larger parameters, our model is very lightweight and has achieved competitive results. Specifically, the parameters of our model are the smallest, our model is only 25.35\% of the parameters of MEG-Net model, but it has achieved a better average PSNR value, which shows the lightweight characteristics of our model.
 In addition, our model has achieved remarkable results on EPFL and INRIA datasets.

 {\bf Qualitative Results.} As shown in Fig.4, regarding qualitative performance, the propose LGFN has proved that it has ability to produce trustworthy details and sharp structures. For SISR methods, VDSR, EDSR and RCAN tends to produce artifacts, and the restored texture details are not clear enough. For LFSR methods, the proposed LGFN has ability to discriminate more dense details. Specifically, in figure ISO\_Chart, the figure recovered by our model is clearer than other figures, and there are fewer artifacts. In the figure Perforated\_Metal\_3, the graph restored by our model has more material texture.

\begin{table*}[htbp] 
\small
\caption{Our team achieved second place on the leader board (last three rows) in the NTIRE-2024 Track 2 Fidelity \& Efficiency test dataset, with quantitative results of 30.05 dB PSNR (average) and 0.924 SSIM (average).}
\centering 
\begin{tabular}{ c |c |c c |c}  
\toprule  
Methods & \#Params & Lytro & Synthetic & Average \\  
\midrule  
Bicubic & — & 25.11 / 0.8404 & 26.46 / 0.8352 & 25.79 / 0.8378 \\  
VDSR \cite{45_v2_32kim2016accurate} & 0.67 M & 27.05 / 0.8888 & 27.94 / 0.8703 & 27.49 / 0.8795 \\  
EDSR \cite{46_v2_33lim2017enhanced} & 38.89 M & 27.54 / 0.8981 & 28.21 / 0.8757 & 27.87 / 0.8869 \\  
RCAN \cite{47_v2_34zhang2018image} & 15.36 M & 27.61 / 0.9001 & 28.31 / 0.8773 & 27.96 / 0.8887 \\  
resLF \cite{20_v2_13zhang2019residual} & 8.65 M & 28.66 / 0.9260 & 29.25 / 0.8968 & 28.95 / 0.9114 \\  
LFSSR \cite{24_v2_17yeung2018light} & 1.77 M & 29.03 / 0.9337 & 29.40 / 0.9008 & 29.21 / 0.9173 \\  
LF-ATO \cite{49_v2_36jin2020light} & 1.36 M & 29.09 / 0.9354 & 29.40 / 0.9012 & 29.24 / 0.9183 \\  
LF-InterNet \cite{25_v2_18wang2020spatial} & 5.48 M & 29.23 / 0.9369 & 29.45 / 0.9028 & 29.34 / 0.9198 \\  
MEG-Net \cite{23_v2_16zhang2021end} & 1.78 M & 29.20 / 0.9369 & 29.54 / 0.9036 & 29.37 / 0.9203 \\  
\hline  
BITSMBU \cite{48_replaceNTIRE2024-LFSR} & 0.66 M & \textbf{\color{red}30.32 / 0.9425} & \textbf{\color{red}30.00 / 0.9095} & \textbf{\color{red}30.16 / 0.9260} \\  
Ours (LGFN-C) & 0.45 M & \textbf{\color{black}30.19 / 0.9402} & \textbf{\color{black}29.92 / 0.9079} & \textbf{\color{black}30.05 / 0.9240} \\  
IIR-Lab \cite{48_replaceNTIRE2024-LFSR} & 0.83 M & 29.96 / 0.9238 & 30.14 / 0.9407 & 29.96 / 0.9238 \\  
\bottomrule  
\end{tabular}    
 
\label{tab:my_table3}  
\end{table*}

\subsection{Ablation Study}

In this section, we further prove the effectiveness of several core parts of the proposed LGFN model through ablation experiments.

{\bf 1) Connection mode of ECAM and ESAM modules.} The connection modes of ECAM and ESAM are classified into cascade connection and parallel connection. In order to verify which connection mode is more effective, we design two models: cascade and parallel, and their corresponding models are LGFN-C and LGFN-P respectively, where LGFN-C is the NTIRE2024 LF image SR competition model. As shown in Table 2, the LGFN-P is better than LGFN-C. The main model of this paper is LGFN-P.

{\bf 2) LGFN w/o DGCE.} The DGCE module is used to extract local feature. To demonstrate the effectiveness of the DGCE module, we remove this module and use parallel ECAM and ESAM modules. As shown in Table 2, the PSNR value is decreased dramatically from 31.8076 dB to 28.7796 dB for 4x SR without DGCE module, and the drop value is 3.028dB. Experiment shows that DGCE module is effective in feature extraction.

{\bf 3) LGFN w/o ESAM.} The ESAM module is used to extract global LF image spatial feature. To demonstrate the effectiveness of the ESAM module, we remove this module. As shown in Table 2, the PSNR value is decreased module, the drop value is 0.1081dB.

{\bf 4) LGFN w/o ECAM.} The ECAM module is used to extract channel feature. To demonstrate the effectiveness of the ECAM module, we remove this module. As shown in Table 2, and the PSNR value is decreased from 31.8076 dB to 31.7804 dB for 4x SR without ECAM module, the drop value is 0.0272dB.

{\bf 5) LGFN w/o ECAM and ESAM.} The ESAM and ECAM module are used to extract global feature. To demonstrate the effectiveness of the ECAM and ESAM modules, we remove these modules. As shown in Table 2, the PSNR value is decreased from 31.8076 dB to 31.6462 dB for 4x SR without them, and the drop value is 0.1614dB, which proves the effectiveness of attention module.

\subsection{NTIRE 2024 LFSR Challenge Results}
The test set of NTIRE2024 LFSR challenge including 16 synthetic LFs and 16 real-world LFs captured by Lytro camera. As shown in Table 3, we proposed a model which ranked the second place in the Track 2 Fidelity \& Efficiency of NTIRE2024 Light Field Super Resolution Challenge with 30.05dB PSNR on the LFSR test dataset.
\section{Conclusion and Feature Work}

In this paper, we investigated the task of lightweight LF image SR and proposed a lightweight LF image SR model named LGFN based on the local similarity and global disparity of SAIs. As a lightweight model, we proposed a feature modulation-based CNN module to extract local features efficiently. Besides, we designed an efficient spatial attention module which uses decomposable large-kernel convolution to enlarge the receptive field and an efficient channel attention module to extract the global features of the LF image. By learning local and global features, our lightweight model has achieved competitive results and ranked the second place in the Track 2 Fidelity \& Efficiency of NTIRE2024 Light Field Super Resolution Challenge and the seventh place in the Track 1 Fidelity.

In our future work, we will adopt model compression techniques, such as knowledge distillation, pruning, and model quantization, to further lighten our model and enhance its effectiveness.

{\bf \large  Acknowledgments}

This work was supported in part by the National Nature Science Foundation of China under Grant No.61901117, No.62301161, A21EKYN00884B03, in part by the Natural Science Foundation of Fujian Province under Grant No.2023J01083.



{
    \small
    \bibliographystyle{unsrt}

}


\end{document}